\documentclass[sigplan,10pt]{acmart}

\renewcommand\footnotetextcopyrightpermission[1]{}
\acmConference[none]{}{}{}

\usepackage{tikz} 
\usepackage{amsmath} 
\usepackage{listings} 
\usepackage{pdfpages} 
\usepackage{soul} 
\usepackage{xspace} 
\usepackage{url} 
\usepackage{graphicx} 
\usepackage{subcaption} 
\usepackage[most]{tcolorbox} 
\usepackage{booktabs} 
\usepackage{multicol} 
\usepackage{wrapfig}
\usepackage{makecell}
\usepackage{enumitem}

\usepackage{minted}

\newcommand{\eat}[1]{} 

\newcommand{\alex}[1]{[{\color{blue}A: #1}]}
\newcommand{\tianle}[1]{[{\color{violet}TL: #1}]}
\newcommand{\ewu}[1]{[{\color{red}wu: #1}]}

\newcommand{\ie}{{\em i.e.}, }

\title{
Toward Systems Foundations for Agentic Exploration
}

\author{Jiakai Xu}
\authornote{Both authors contributed equally to this research.}
\affiliation{%
\institution{Columbia University}
\streetaddress{}
\city{}
\state{}
\country{}
\postcode{}
}
\email{ax2155@columbia.edu}
\orcid{0009-0006-2074-9109}

\author{Tianle Zhou}
\authornotemark[1]
\affiliation{%
\institution{Columbia University}
\streetaddress{}
\city{}
\state{}
\country{}
\postcode{}
}
\email{mz2998@columbia.edu}

\author{Eugene Wu}
\affiliation{%
\institution{Columbia University}
\streetaddress{}
\city{}
\state{}
\country{}
\postcode{}
}
\email{ewu@cs.columbia.edu}
\orcid{0000-0003-4254-6688}

\author{Kostis Kaffes}
\affiliation{%
\institution{Columbia University}
\streetaddress{}
\city{}
\state{}
\country{}
\postcode{}
}
\email{kkaffes@cs.columbia.edu}
\orcid{0000-0002-0517-7206}
\begin{document}

\begin{abstract}
Agentic exploration, letting LLM-powered agents branch, backtrack, and search across many execution paths, demands systems support well beyond today’s pass-@-k resets. Our benchmark of six snapshot/restore mechanisms shows that generic tools such as CRIU or container commits are not fast enough even in isolated testbeds, and they crumble entirely in real deployments where agents share files, sockets, and cloud APIs with other agents and human users. In this talk, we pinpoint three open fundamental challenges: fork semantics, which concerns how branches reveal or hide tentative updates; external side-effects, where fork awareness must be added to services or their calls intercepted; and native forking, which requires cloning databases and runtimes in microseconds without bulk copying.
\end{abstract}

\settopmatter{printfolios=true}

\maketitle

\pagestyle{plain}


\section{Agentic Exploration}
\label{sec:exploration}


Large Language Models (LLMs) under an interaction–feedback paradigm have demonstrated strong performance on everyday tasks, where prior interactions determine following actions \citep{zheng2023judgingllmasajudgemtbenchchatbot,alpaca_eval}.
More recently, LLM-powered agents functioning as system agents are used to interact directly with real computing environments, such as operating systems and development toolkits~\citep{xie2024osworldbenchmarkingmultimodalagents,zhou2023webarena,eliseeva2025envbenchbenchmarkautomatedenvironment,jimenez2024swebench,tbench_2025}. 
These tasks often require actions that can alter the state, e.g., an application, a database, a language runtime, or an operating system, making the problem a partially observable, multi-step decision process.
Consequently, effective exploration—where an agent actively interacts with a stateful environment, observes the outcome of its actions, and making better multi-step strategies—becomes critical.
On Terminal-Bench's~\citep{tbench_2025} command-line tasks, disabling exploration reduces accuracy by 27.2 percentage points (30.6\% $\rightarrow$ 3.4\%).

\paragraph{From pass@k to real exploration}
Most exploration-based agent frameworks implicitly assume that the environment can be restored to an initial reference state such that applying the same set of actions leads to identical observations \citep{towers2024gymnasiumstandardinterfacereinforcement,openaigym}.
This guarantees that exploration outcomes on alternative branches are valid and reproducible.
In practice, this is satisfied by benchmark harnesses that deliver deterministic initial states and allow programmatic resets.
For example, WebArena\citep{zhou2023webarena} constructs each website as a self-contained Docker image and provides scripts to reset to the initial state, and OSWorld\citep{xie2024osworldbenchmarkingmultimodalagents} offers task-specific initial-state setup and uses VM images to recover the initial state.
In these scenarios, the baseline is always to pass@k from a clean state: for each trial the harness resets to a pristine snapshot and allows the agent to act until success/failure.

While the pass@k method is simple and works well when per-step overheads are small or tasks are short, it performs poorly on more complex, long-horizon, and realistic tasks because agents tend to make mistakes due to losing sight of ultimate goals and cumulative errors, i.e., non-observable state changes, on long-horizon tasks~\citep{erdogan2025planandactimprovingplanningagents}.
The most common solution to this problem is taking into account intermediate states and doing exploration over them.
For instance, Reflective Monte Carlo Tree Search (MCTS)~\citep{yu2025exactteachingaiagents} augments tree search with budgeted rollouts and reward backtracking, and HiAgent\citep{hu2024hiagenthierarchicalworkingmemory} uses hierarchical decomposition with task-level checkpointing, both obtained impressive performance gains in long-horizon tasks.
On terminal-bench, simply allowing for searching from intermediate states significantly improves agent performance:
We observe a 20 percentage point increase in success rate on a selected subset of tasks when applying  MCTS instead of the baseline pass@2 method with claude 3.5 sonnet model.


\section{Exploration as State Restoration}
\begin{figure}[htbp]\raggedright
  \includegraphics[width=1\linewidth]{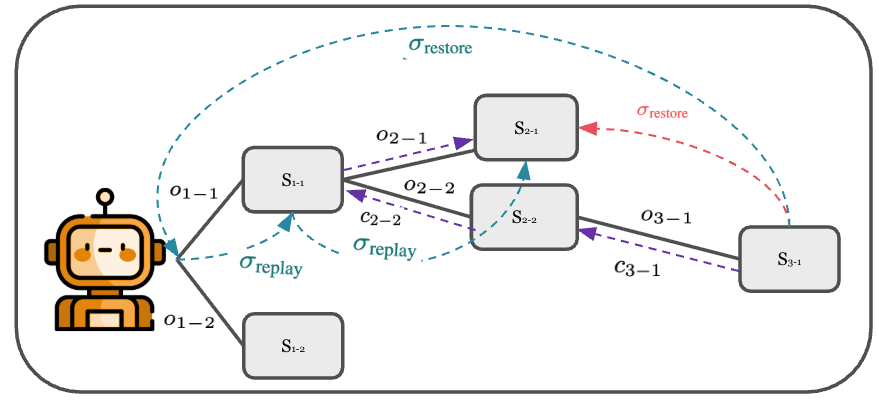}
  \caption{An LLM agent (orange) explores by taking different actions, creating a branching tree in which every node represents a distinct state of the environment. A prefix replay (teal) starts from root and replay all commands on record; A snapshot/restore (red) checkpoint method restores the state directly to the target; a backtracking method (purple) goes through all intermediate nodes to the target node. \eat{\ewu{This figure is'nt referenced in the text, and it's not clear what its point is.  What about the arrows are we trying to emphasize with an example?   If you choose to keep the figure, please address the feedback on this diagram I gave on slack in the past.    the red and blue nodes and their shades don't mean anything so get rid of them.   arrows and labels should be legible and explained.  For instance, why does the replace arrows go from  Si-1 to S2-1?.   Simplify the tree to a minimal example and construct a real example.}}}
  \label{fig:exploration-pipeline}
\end{figure}

\eat{\tianle{Checkpoint is for go from a intermediate state and start from it. You can undo -- but for parallel situation, it will fail. So doing snapshot is a fundamental method for doing exploration. Thus, checkpoint/restore should be a primitive.} \alex{Yes, the parallelism is one point, although we seems haven't test parallel exploration, our framework should technically support that.} \tianle{In real world, there will be different types of states to be recorded. There will be emphmoral states, static states such as fs. Different types of tasks require modfiy and resume different types of states. Current snapshot are static - however, different tasks have different scopes. There will be different copes -> its tricky becau}se getting right enough to be recorded.}
Instead of always resetting to a clean initial state, supporting exploration from intermediate states requires digital environments to resume execution from that point onward.  The system infrastructure must ensure that replaying or branching from this state produces observations consistent with the original execution. Systems can support such agentic exploration with three different primitives (as displayed in Figure\ref{fig:exploration-pipeline}):

(1) \emph{Replay-to-node (prefix replay)}. For every explored search-tree node, the runtime records the command prefix needed to reach the node from the initial state.
Revisiting the node is achieved by re-executing the command prefix. This obviates any explicit state capture overhead, but incurs replay cost proportional to the length of the command prefix and the overhead of the individual commands.

(2) \emph{Snapshot/Restore.} Alternatively, the system can materialize a snapshot at each node and reload it on demand, trading storage overhead for $O(1)$ restoration latency.

(3) \emph{Backtracking.} For any operation $o_i$ made that shifts the environment state $S_i\to S_i'$, a compensation operation $c_i=reverse(o_i)$ is pre-defined that shifts $S_i'\to S_i$. 
Restoration amounts to reversing all of the intermediate nodes, but relies on pre-defined logic~\citep{chang2025sagallmcontextmanagementvalidation}. 

Unlike humans, agent exploration involves frequent state switching, which requires high-fidelity state restoration. Backtracking method, in practice, is therefore challenging because many system-level operations are inherently irreversible 
(e.g., file deletion, network I/O, time-sensitive actions). Therefore, the backtracking method is not good for general-purpose exploration. Thus, a reliable agent system framework must at least provide a minimal form of snapshot/restoration, ensuring that distinct exploration operations can be conducted with consistent observations.

In the simplest agent settings---such as a dialog agent whose entire world state is the conversation log---snapshotting reduces to persisting that log, so recording the full state is nearly trivial.
In richer cases where the agent manipulates a stateful environment, e.g., external software or operating-system resources, snapshotting must encompass the full execution context. At minimum, this includes: 

\begin{itemize}[leftmargin=*]
\item \textbf{Filesystem} — to preserve file modifications, e.g., installed packages and intermediate artifacts in long-running tasks.
\item \textbf{Memory} — to retain application and kernel state, e.g., heap memory.
\end{itemize}

Thus, a full system snapshot is necessary to enable seamless, multi-path exploration in realistic tasks.
This insight motivates the benchmark study that follows.

\subsection{Existing Technologies and Benchmark}

\begin{table*}[ht]
    \centering
    \caption{Snapshot and restore time (in seconds) across different tools and configurations.}
    \label{tab:benchmark}
    \begin{tabular}{|c|c|c|c|c|c|c|c|c|}
        \hline
        \textbf{Operation} & \textbf{Memory} & \textbf{Disk} & \textbf{criu} & \textbf{Docker} & \textbf{Podman} & \textbf{checkpoint-lite} & \textbf{Hybrid} & \textbf{AWS-VM} \\
        \hline
        Snapshot + Restore & 0 GB & 0 GB & 0.060 & 0.416 & 0.835 & 0.418 & 1.657 & 353 \\
        Snapshot + Restore & 1 GB & 0 GB & 0.760 & / & / & 1.079 & 9.921 & - \\
        Snapshot + Restore & 2 GB & 0 GB & 1.445 & / & / & 1.757 & 18.154 & - \\
        Snapshot + Restore & 0 GB & 1 GB & / & 5.097 & 7.935 & 2.499 & 14.735 & - \\
        Snapshot + Restore & 0 GB & 2 GB & / & 6.915 & 12.914 & 4.622 & 26.648 & - \\
        \hline
    \end{tabular}
    \vspace{0.5ex}\newline
    {\footnotesize 
    Tested on a Linux server with 56-core Intel® Xeon® Gold 5512U CPU, 128GB RAM, running Ubuntu 24.04.2 LTS with Linux kernel 6.8.0.  \\
    Tool versions: CRIU 4.1, Docker 27.5.1, Podman 4.9.3, runc 1.2.5. checkpoint-lite is our own Go-based tool using CRIU + OverlayFS.}
\end{table*}


We measured snapshot and restore latency as the amount of modified application memory or filesystem contents are independently varied from $0GB$ to $2GB$.  We compared six commonly-used mechanisms: \textbf{CRIU}\cite{criu}, \textbf{Docker}, \textbf{Podman}, \textbf{Hybrid} (Podman + CRIU checkpoint), \textbf{AWS VM snapshots}, and our prototype \textbf{checkpoint-lite} (CRIU + OverlayFS).
Table~\ref{tab:benchmark} summarizes the results, and five trends stand out:

\begin{enumerate}[leftmargin=*]
\item \textbf{AWS VMs}, used by OS-World~\cite{xie2024osworldbenchmarkingmultimodalagents}, are extremely slow to re-instantiate as they are not built for this purpose. 
\item \textbf{Docker/Podman commits}, used by WebArena \citep{zhou2023webarena}, AgentBench \citep{liu2023agentbenchevaluatingllmsagents}, and Terminal-Bench \citep{tbench_2025}, rebuild containers from image layers, \ie filesystem state only and therefore lose live memory. Startup latencies can exceed 10 s, making them unsuitable for fine-grained agentic exploration.
\item \textbf{CRIU} offers fast memory snapshots by dumping process memory and metadata to a file, snapshotting and restoring 2 GiB process in 1.445 s, but snapshot cost rises linearly with memory and it still ignores files.
\item \textbf{Hybrid (Podman checkpoint)} integrates CRIU with container runtimes to capture memory and network, but restore times remain high (up to 12 s for 2 GiB).
\item Our Go-based \textbf{checkpoint-lite} prototype orchestrates CRIU dumps alongside OverlayFS layer snapshots, achieving near-CRIU times (1.757 s for 2 GiB state) while also preserving filesystem state.
\end{enumerate}

\noindent However, even before factoring in storage costs, existing checkpoint/restore tools impose second-scale overheads, making them unsuitable for rapid agentic exploration.
Worse, they are missing critical features that we show next are essential for environment-agnostic agentic exploration.

\section{The Missing Pieces}


\paragraph{From snapshot/restore to native forking}

What agentic exploration really demands is not generic \emph{snapshot/restore} but a lightweight, \emph{native fork} primitive: the ability to spin off multiple live logical copies of a running application or system without duplicating unchanged data.
Conceptually, it resembles \texttt{fork()} in Unix---copy-on-write pages, lazy duplication---but extended to encompass higher-level resources.
Unlike traditional OS forks, an agent-targeted fork must duplicate open file descriptors \emph{semantically}: a child’s write to a socket should not corrupt the parent’s stream, and diverging file writes should land in per-branch overlays that can later merge or discard cleanly.
Achieving this requires tighter integration between the OS, the storage stack, and the language runtime so that forking incurs microseconds of latency rather than the milliseconds or seconds we observe with coarse snapshotting.

Generic, system-wide forking is useful, but some subsystems benefit from domain-specific support.
Databases are a prime example: Neon’s “branching” Postgres \citep{neon_github_software} clones let developers fork a live logical database, yet each branch takes seconds to materialize--—far slower than the sub-millisecond forks agents would need for interactive branching.
Similar gaps appear in language runtimes: Python’s multiprocessing fork inherits bytecode and heap, but extension modules holding GPU tensors or open sockets do not survive, forcing full re-initialization. 
Bridging this gap calls for \emph{native fork hooks} inside components—--database engines that version page caches in micro-seconds and runtimes that expose copy-on-write heaps.
Building such primitives pushes the responsibility down to where the semantics are understood, yielding fork operations that are both correct and fast enough to unlock large-scale, multi-path agentic exploration.

\paragraph{From benchmarks to the real world.}
The snapshot mechanisms described above suffice only for \emph{isolated} benchmarking environments.
Real deployments couple agents to databases, browsers, and cloud APIs whose state lives beyond the local filesystem or RAM.
The simplest example is a live socket: restoring a checkpoint invalidates the TCP sequence numbers, auth tokens, or DOM tree held by the remote peer.
Thus, we need to develop methods to enable general-purpose agentic exploration without sacrificing correctness at scale.
One such example could be to expose \emph{fork-aware APIs} whose side effects are intrinsically versioned—much like S3’s object-versioning, where each branch writes to an immutable commit rather than mutating shared state.

\paragraph{Semantics of multi-agent exploration}
In production settings, multiple autonomous agents---and often live human users---operate on the same resources at once, so the key semantic question is what those other actors should observe while one agent is branching speculatively.
A conservative design might reveal only \emph{committed} trajectories, hiding tentative side effects until they are finalized; this preserves serial consistency but can cause costly merge conflicts.
A more optimistic design could let agents fork atop one another's in-flight trajectories, promoting richer collaboration yet exploding the state space combinatorially.

\begin{acks}
This work was supported by the National Science Foundation (1845638, 1740305, 2008295, 2106197, 2103794, 2312991) and DAPLab funders (Amazon, Google, Intellect Design, Tidalwave, Veris).
\end{acks}

\bibliographystyle{ACM-reference-format}
\bibliography{main.bib}

\end{document}